\documentclass[10pt,twocolumn,american,aps,manuscript,10pt,a4paper,twocolumn,american,twocolumn,prl,a4paper,footinbib]{revtex4-1}
\usepackage{lmodern}
\usepackage[T1]{fontenc}
\usepackage[latin9]{inputenc}
\usepackage[a4paper]{geometry}
\usepackage{float}
\usepackage{amsmath}
\usepackage{amssymb}
\usepackage{graphicx}
\PassOptionsToPackage{normalem}{ulem}
\usepackage{ulem}

\makeatletter
\@ifundefined{textcolor}{}
{%
 \definecolor{BLACK}{gray}{0}
 \definecolor{WHITE}{gray}{1}
 \definecolor{RED}{rgb}{1,0,0}
 \definecolor{GREEN}{rgb}{0,1,0}
 \definecolor{BLUE}{rgb}{0,0,1}
 \definecolor{CYAN}{cmyk}{1,0,0,0}
 \definecolor{MAGENTA}{cmyk}{0,1,0,0}
 \definecolor{YELLOW}{cmyk}{0,0,1,0}
}

\makeatother

\usepackage{babel}
\begin{document}
\title{Strong disorder in nodal semimetals: Schwinger-Dyson\textendash Ward
approach}
\author{Björn Sbierski and Christian Fräßdorf}
\affiliation{Dahlem Center for Complex Quantum Systems and Institut für Theoretische
Physik, Freie Universität Berlin, D-14195, Berlin, Germany}
\begin{abstract}
The self-consistent Born approximation quantitatively fails to capture
disorder effects in semimetals. We present an alternative, simple-to-use
non-perturbative approach to calculate the disorder induced self-energy.
It requires a sufficient broadening of the quasiparticle pole and
the solution of a differential equation on the imaginary frequency
axis. We demonstrate the performance of our method for various paradigmatic
semimetal Hamiltonians and compare our results to exact numerical
reference data. For intermediate and strong disorder, our approach
yields quantitatively correct momentum resolved results. It is thus
complementary to existing RG treatments of weak disorder in semimetals. 
\end{abstract}
\date{\today}

\maketitle
\emph{Introduction.}\textemdash Semimetals with point-like Fermi surface
are by now an established research field in condensed matter physics.
Well-studied examples are two-dimensional (2d) Dirac fermions in graphene
\cite{Novoselov2005}, 3d Weyl fermions in spin-orbit coupled compounds
\cite{Bernevig2015}, or parabolic band touching points in Bernal-stacked
bilayer graphene \cite{McCann2013}. Many experimental properties
of semimetals rely on the presence of impurities or disorder, ubiquitous
in solid state realizations, but under control in cold atom setups
via speckle potentials \cite{Volchkov2018}. For example, in undoped
graphene the non-zero density-of-states is purely disorder generated
\cite{DasSarma2011}. Likewise, in ARPES experiments it is the disorder,
which broadens the spectral function at the nodal point and modifies
its dispersion away from it. Another example is a disorder driven
phase transition between a semimetallic and a metallic phase in 3d
Weyl fermions \cite{Syzranov2016c}. In the following, motivated by
the observables described above, we focus on single-particle properties.

Theoretically, however, the currently available analytical methods
for disordered semimetals yield unsatisfactory results. Weak disorder
in semimetals can be treated using the perturbative Wilsonian momentum
shell renormalization group (RG) as pioneered in the context of 2d
Dirac systems by Ref. \cite{Ludwig1994}. The starting point of this
approach is the functional integral formalism that can be disorder
averaged after the fermions have been replicated. The bosonic disorder
field is then eliminated in favor of a four-fermion pseudo-interaction
whose coupling constant flows as high energy shells are integrated
out perturbatively. The drawback of the perturbative RG method is
its inapplicability in the strong disorder regime and the difficulty
to extract quantitative information about observables from the abstract
RG flow. Another standard approach to disorder is the non-perturbative
self-consistent Born approximation (SCBA). For metals, its validity
relies on the smallness of the parameter $1/k_{F}l$ that quantifies
the suppression of diagrams with crossed disorder lines omitted in
SCBA \cite{BruusBook}. Here, $k_{F}$ is the Fermi momentum and $l$
the mean free path. However, for semimetals with $k_{F}\!=\!0$, the
SCBA cannot be justified \cite{Ostrovsky2006,Aleiner2006}.

In this work, we propose a novel non-perturbative approach to disorder,
systematically going beyond the SCBA but free of its above restrictions.
Our approach is applicable for strong and intermediate Gaussian disorder
where it yields a quantitatively accurate and momentum dependent self-energy.
For semimetals, it is thus complementary to the RG approach. We start
from the Fermi-Bose field theory mentioned above, but we do not integrate
out the disorder field. Instead we derive an exact Schwinger-Dyson
equation \cite{Peskin1995,Kopietz2010} for the fermionic self-energy,
which is closed by replacing the Fermi-Bose three-vertex with the
help of a Ward-identity. This replacement is justified for strong
disorder only. We arrive at a set of ordinary differential equations
on the imaginary frequency axis that can be easily solved numerically.
We apply this Schwinger-Dyson\textendash Ward (SDW) approach for various
paradigmatic semimetals and compare our results to exact numerical
reference data computed with a dedicated momentum space version of
the kernel polynomial method. We also compare our results to the SCBA
and a semi-classical approximation for strong disorder.

\emph{Model and main result.}\textemdash We consider a generic two-band
semimetal Hamiltonian $H_{0}(\vec{k})$ with a degeneracy point at
$\vec{k}=0$ located at zero energy, $H_{0}(\vec{k}\!=\!0)\!=\!0$.
For simplicity, we assume an isotropic dispersion $\pm E_{0}(k)$
with particle-hole symmetry. These assumptions are not crucial in
the following but valid for many popular choices of $H_{0}(\vec{k})$
like Dirac nodes.

We add a smoothly correlated disorder potential $V(\vec{r})$ which
is assumed to be diagonal in band space. Its correlation length $\xi$
represents the disorder puddle's typical linear extent. We define
the fundamental energy scale $E_{\xi}\!=\!E_{0}(k\!=\!1/\xi)$. For
the statistical properties of $V$, we assert a Gaussian probability
distribution $P\left[V\right]$ and define the disorder average of
a quantity $Q[V]$ as $\left\langle Q\right\rangle _{dis}=\int\mathcal{D}VQ[V]P\!\left[V\right]$.
We assume the disorder correlator $\mathcal{K}(\vec{r}-\vec{r}^{\prime})$
to have a Gaussian shape 
\begin{equation}
\mathcal{K}(\vec{r}-\vec{r}^{\prime})=\left\langle V(\vec{r})V(\vec{r}^{\prime})\right\rangle _{dis}\!=\!\frac{W^{2}E_{\xi}^{2}}{(2\pi)^{d/2}}e^{-\frac{1}{2}|\vec{r}-\vec{r}^{\prime}|^{2}/\xi^{2}}.\label{eq:K}
\end{equation}
The dimensionless parameter $W$ quantifies the strength of disorder.
It relates the typical potential in a puddle $\sim\sqrt{\langle V(\vec{r})^{2}\rangle_{dis}}$
to the kinetic energy of a particle confined to the puddle's volume.
We refer to $W\ll1$ as weak and $W\gg1$ as strong disorder, respectively.
We are interested in the zero-temperature retarded Green function
$G_{V}^{R}(\omega)=(\omega+i\eta-H_{0}-V)^{-1}$ and, in particular,
in its disorder average $\langle G_{V}^{R}\rangle_{dis}\equiv G^{R}$
,
\begin{align}
G^{R}(\omega,\vec{k}) & =\frac{1}{\omega+i\eta-H_{0}(\vec{k})-\Sigma^{R}(\vec{k},\omega)},\label{eq:Sigma_def}
\end{align}
at the nodal point energy $\omega=0$. This defines the disorder induced
self-energy $\Sigma^{R}(\vec{k},\omega)$. Our main result is a self-consistency
equation for the self-energy on the imaginary frequency axis,
\begin{align}
 & \Sigma_{\sigma_{1}\sigma_{2}}(i\omega,\vec{k})=\label{eq:SD-Ward}\\
 & \sum_{\sigma}\int_{\vec{q}}\mathcal{K}(\vec{q})\left[\delta_{\sigma_{1}\sigma}-\partial_{i\omega}\Sigma_{\sigma_{1}\sigma}(i\omega,\vec{k})\right]G_{\sigma\sigma_{2}}(i\omega,\vec{k}+\vec{q}),\nonumber 
\end{align}
from which one may calculate $\Sigma^{R}(\omega\!=\!0,\vec{k})$.
The structure of Eq. (\ref{eq:SD-Ward}) is reminiscent of the SCBA
with $\partial_{i\omega}\Sigma$ as a correction term. The derivation
of Eq. (\ref{eq:SD-Ward}), which relies on Schwinger-Dyson equations,
a Ward-identity and an approximation that is valid for a sufficient
broadening of the quasiparticle pole, will be sketched after treating
a few examples. We refer to Eq. (\ref{eq:SD-Ward}) as the Schwinger-Dyson\textendash Ward
approximation (SDWA).

To solve Eq. (\ref{eq:SD-Ward}), we parameterize the momentum dependence
of $\Sigma(i\omega,\vec{k})$ using the symmetries of the clean Hamiltonian
$H_{0}$, which are restored after the disorder average. We isolate
the derivative term and discretize the momentum dependence, which
yields a system of first order ordinary differential equations (ODE)
in $\omega$ \footnote{The resulting equation has a similar mathematical structure as a self-energy
flow equation in a functional RG approach.}. For the boundary condition, $\underset{\omega\rightarrow\infty}{\mathrm{lim}}G(i\omega)=\frac{1}{i\omega}$
\cite{Negele1988} implies an at most sublinear asymptotic of $\Sigma\left(i\omega\right)$
in $i\omega$. Hence, at $\omega=\omega_{max}\gg E_{\xi}$, we can
approximately neglect $\partial_{i\omega}\Sigma(i\omega)$ in Eq.
(\ref{eq:SD-Ward}). The resulting self-consistent solution for $\Sigma(i\omega_{max})$
can be found algebraically in the limit $\omega_{max}\gg E_{\xi}$.
We finally apply standard routines to solve the array-valued ODE numerically.
We have checked that the results for $\Sigma(i\omega=i\eta)$ do not
depend on (large enough) $\omega_{max}$.

\emph{Exact numerical results from KPM.}\textemdash To gauge the quality
of the SDW approximation, we employ the kernel polynomial method (KPM)
\cite{Weisse2006} to obtain numerically exact reference data for
$\Sigma_{\sigma_{1}\sigma_{2}}^{R}(\omega,\vec{k})$. The standard
iteration procedure of KPM repeatedly applies the Hamiltonian as $(H_{0}+V)\left|\psi\right\rangle $.
In contrast to recent state-of-the art studies for disordered Weyl
nodes \cite{Pixley2017}, we work in momentum space $\left|\psi\right\rangle =\sum_{\vec{k}\sigma}\psi_{\vec{k}\sigma}|\vec{k}\sigma\rangle$,
thus avoiding to regularize $H_{0}$ on a lattice. While $H_{0}$
is diagonal in $\vec{k}$, the potential $V$ is diagonal in real
space. We employ a fast Fourier transform (FFT) on $\psi_{\vec{k}\sigma}$
to get to real space, apply $V$ and transform back to momentum space.
We use an equidistant $k$-space grid with spacing $\Delta k\ll\xi^{-1}$
and a UV-cutoff $\text{\ensuremath{\Lambda\gg\xi^{-1}}}$. We checked
the convergence of our final results with respect to the number of
moments, disorder realizations and lattice points. The limitation
of the KPM is in the weak disorder regime, when the finite-size energy
starts to compete with the disorder induced energy scale.
\begin{figure}
\noindent \begin{centering}
\includegraphics{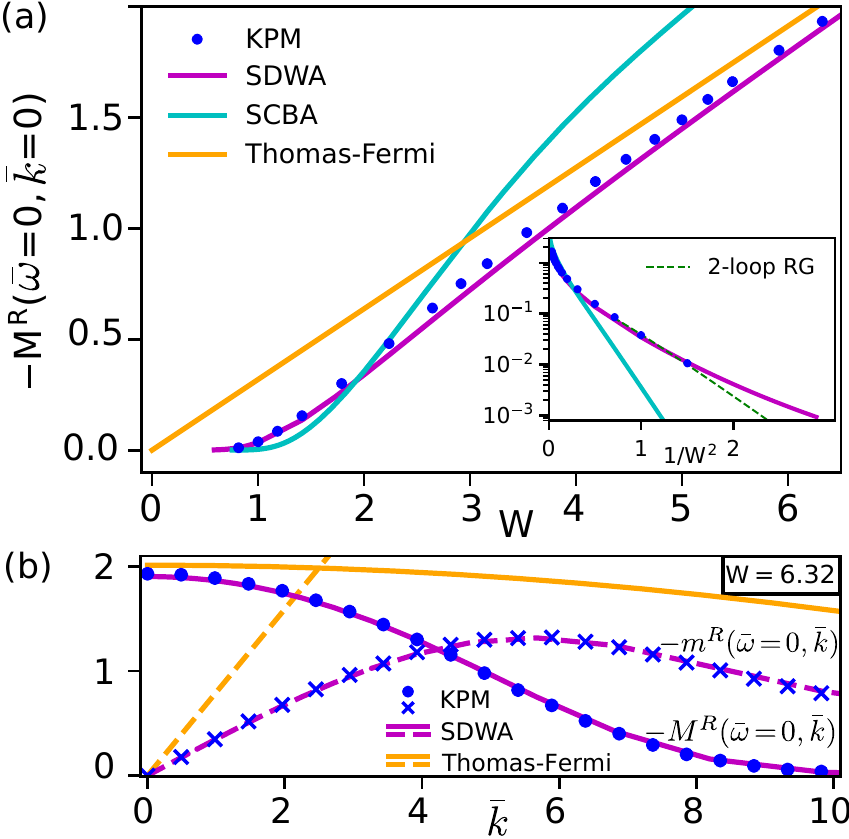}
\par\end{centering}
\caption{{\small{}\label{fig:res_2dDirac}(a) Imaginary part of the disorder
induced retarded self-energy for a 2d Dirac node at the nodal point
as a function of disorder strength $W$. Our }SDWA{\small{} result
(magenta line) compares well to the exact KPM data (blue dots) and
asymptotically agrees with the Thomas-Fermi approximation (orange).
The SCBA result is shown in cyan. Inset: For very small disorder,
the SDWA deviates from the scaling found by RG (green dashed line).
(b) The momentum dependence of the self-energy at $W=6.32$ calculated
from KPM (blue symbols), the SDWA (magenta lines) and Thomas-Fermi
approximation (orange lines). }}
\end{figure}

\emph{Application to 2d Dirac node.}\textemdash We now apply the SDWA
to the case of a disordered 2d Dirac node, $H_{0}(\vec{k})=\hbar v(\sigma_{x}k_{x}+\sigma_{y}k_{y})$,
with $E_{0}(k)=\hbar vk$ and the fundamental energy scale $E_{\xi}\!=\!\hbar v/\xi$.
Using dimensionless variables $\vec{\bar{k}}\!=\!\vec{k}/\xi$ and
$\bar{\omega}\!=\!\omega/E_{\xi}$, the ansatz for the self-energy
reads 
\begin{equation}
\Sigma(i\bar{\omega},\vec{\bar{k}})/E_{\xi}=m(\bar{\omega},\bar{k})[\sigma_{x}\cos\bar{\phi}+\sigma_{y}\sin\bar{\phi}]+iM(\bar{\omega},\bar{k}),\label{eq:Sigma_2dDirac}
\end{equation}
where we switched to polar coordinates for $\vec{\bar{k}}$ on the
right hand side. While $m$ represents a renormalization of $H_{0}$,
$M$ can be interpreted as a scattering rate. Using this ansatz in
Eq. (\ref{eq:SD-Ward}), we obtain two coupled ODEs: 
\begin{align}
\partial_{\bar{\omega}}M(\bar{\omega},\bar{k}) & =1\!+\!\frac{J_{0}(\bar{\omega},\bar{k})M(\bar{\omega},\bar{k})\!+\!J_{1}(\bar{\omega},\bar{k})m(\bar{\omega},\bar{k})}{W^{2}[J_{0}^{2}(\bar{\omega},\bar{k})+J_{1}^{2}(\bar{\omega},\bar{k})]},\label{eq:DM}\\
\partial_{\bar{\omega}}m(\bar{\omega},\bar{k}) & =\frac{J_{0}(\bar{\omega},\bar{k})m(\bar{\omega},\bar{k})-J_{1}(\bar{\omega},\bar{k})M(\bar{\omega},\bar{k})}{W^{2}[J_{0}^{2}(\bar{\omega},\bar{k})+J_{1}^{2}(\bar{\omega},\bar{k})]},\label{eq:Dm}
\end{align}
where the functions $J_{0}$ and $J_{1}$ themselves depend on $m$
and $M$ as
\begin{equation}
\left\{ \begin{array}{c}
\!\!\!J_{0}(\bar{\omega},\bar{k})\!\!\!\\
\!\!\!J_{1}(\bar{\omega},\bar{k})\!\!\!
\end{array}\right\} \!=\!\!\int_{0}^{\infty}\!\!\!\!d\bar{q}\left\{ \begin{array}{c}
\!\!\!\tilde{M}(\bar{\omega},\bar{q})I_{0}(\bar{k}\bar{q})\!\!\!\\
\!\!\!\tilde{m}(\bar{\omega},\bar{q})I_{1}(\bar{k}\bar{q})\!\!\!
\end{array}\right\} \!\!\frac{\bar{q}e^{-(\bar{q}^{2}+\bar{k}^{2})/2}/(2\pi)}{\tilde{m}^{2}(\bar{\omega},\bar{q})\!+\!\tilde{M}^{2}(\bar{\omega},\bar{q})}.
\end{equation}
Here, $I_{j}$ are modified Bessel functions of the first kind \cite{Gradshteyn2007},
$\tilde{m}(\bar{\omega},\bar{k})\!=\!\bar{k}\!+\!m(\bar{\omega},\bar{k})$
and $\tilde{M}(\bar{\omega},\bar{k})\!=\!\bar{\omega}-\!M(\bar{\omega},\bar{k})$.
The initial conditions for $\bar{\omega}_{max}\gg1$ read $M(\bar{\omega}_{max},\bar{k})=-\frac{W^{2}}{2\pi\bar{\omega}_{max}}$
and $m(\bar{\omega}_{max},\bar{k})=0$.

The set of ODEs (\ref{eq:DM}) and (\ref{eq:Dm}) can be solved numerically
after discretizing the $\bar{k}$-dependence of $m$ and $M$ on a
geometric grid. In Fig. \ref{fig:res_2dDirac}(a) we compare the resulting
disorder induced self-energy at the pole of the clean Green function,
$M^{R}(\bar{\omega}=0,\bar{k}=0)=M(i\bar{\omega}=i\eta,\bar{k}=0)$
(magenta line) to the exact KPM data (blue dots), finding good agreement.
This is true even for the smallest disorder strength $W\simeq0.8$
that we can reach with KPM, see inset. Based on the above comment
about the validity of SDWA, we consider this agreement for $M^{R}(\bar{\omega}=0,\bar{k}=0)\ll1$
as coincidental. In fact the SDW solution for $M^{R}(\bar{\omega}=0,\bar{k}=0)$
does not agree asymptotically with the form $\sim\exp(-\pi/W^{2})/W$,
that is inferred from the scale where the 2-loop Wilsonian RG-flow
crosses over to strong disorder \cite{Schuessler2009} (green dashed
line in the inset). In the supplemental material \footnote{See Supplemental Material at
[URL will be inserted by publisher] for a discussion of weak disorder in a 2d Dirac node, the Thomas-Fermi approximation for strong disorder, details on the SDWA for 3d Weyl and parabolic 2d semimetal and a derivation of the Schwinger-Dyson-Ward approximation} we show additional
KPM data confirming the validity of the RG result for weak disorder,
albeit using a modified uncorrelated disorder model, where even smaller
$M^{R}$ can be resolved. 

In Fig. \ref{fig:res_2dDirac}(a), we also illustrate the failure
of the SCBA for all disorder strengths \cite{Aleiner2006,Ostrovsky2006}
(cyan line). The data is obtained from an iterative numerical solution
of the SCBA-equation, i.e. Eq. (\ref{eq:SD-Ward}) without the derivative.

The momentum dependence of the self-energy at $W=6.32$ is addressed
in Fig. \ref{fig:res_2dDirac}(b). Again, the SDW results (dashed
and solid magenta lines) compare well with exact KPM data (blue symbols).
Note that $m^{R}(\bar{\omega}=0,\bar{k})\propto-\bar{k}$ encodes
a velocity suppression.

In the limit of large disorder, $W\gg1$, the typical electron wavelength
(at zero total energy) is on the order of $\xi/W$, thus the electron
motion in the disorder potential varying on the scale $\xi$ can be
approximated as semi-classical \cite{Shklovskii1984,Skinner2014,Pesin2015}.
This motivates the Thomas-Fermi approximation, $G^{R}(\omega,\vec{k})=\int_{-\infty}^{\infty}dU\,P_{1}(U)[\omega+i\eta-H_{0}(\vec{k})-U]^{-1}$
where $P_{1}(U)=\exp\left(-\frac{U^{2}}{2\mathcal{K}(\vec{r}=0)}\right)/\sqrt{2\pi\mathcal{K}(\vec{r}=0)}$
is the distribution function of the disorder potential at a single
point (see supplement \cite{Note2} for details). At the nodal point, the Thomas-Fermi
estimate \cite{Trappe2015,Volchkov2018} $M^{R}(\bar{\omega}\!=\!0,\bar{k}\!=\!0)E_{\xi}=-\frac{1}{\pi P_{1}(0)}$
agrees with the KPM and SDWA asymptotically {[}orange line in Fig.
\ref{fig:res_2dDirac}(a){]}. Consequently, this result can also be
reproduced analytically from Eq. (\ref{eq:SD-Ward}) after setting
$H_{0}\rightarrow0$. However, for finite momentum, the Thomas-Fermi
approximation fails even for strong disorder, see Fig. \ref{fig:res_2dDirac}(b).
\begin{figure}
\noindent \begin{centering}
\includegraphics{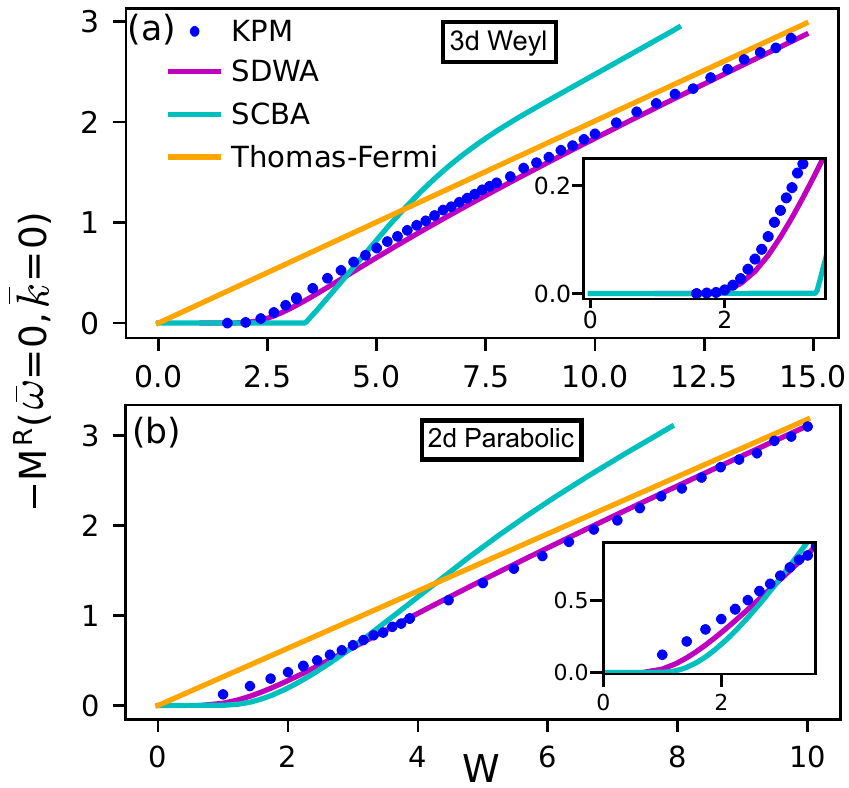}
\par\end{centering}
\caption{{\small{}\label{fig:res_3dWeyl_2dParabolic}The same as in Fig. \ref{fig:res_2dDirac}(a)
but for (a) a 3d Weyl and (b) a 2d parabolic semimetal. The }SDW{\small{}
results (magenta) compare well to the exact KPM data (blue) except
for small $|M^{R}(\bar{\omega}=0,\bar{k}=0)|$, see insets.}}
\end{figure}

\emph{Other dispersions.}\textemdash To show the flexibility of the
SDWA, we now modify the clean Hamiltonian $H_{0}$, modeling other
types of nodal semimetals. First, we consider the case of a 3d Weyl
node, $H_{0}(\vec{k})=\hbar v(\sigma_{x}k_{x}+\sigma_{y}k_{y}+\sigma_{z}k_{z})$,
with $E_{0}$ and $E_{\xi}$ as before. The Weyl node features a disorder
induced phase transition \cite{Fradkin1986,Syzranov2016c}, for $W$
below a critical disorder strength $W_{c}$, the self-energy vanishes
at $k\!=\!0$. The ansatz for the self-energy and the resulting modifications
to the ODEs (\ref{eq:DM}) and (\ref{eq:Dm}) are detailed in the
supplement \cite{Note2}. Fig. \ref{fig:res_3dWeyl_2dParabolic}(a) compares the
SDW results for $M^{R}(\bar{\omega}\!=\!0,\bar{k}\!=\!0)$ to KPM,
SCBA and the Thomas-Fermi approximation. Again, while the SCBA fails
quantitatively, our SDWA is in good agreement with the exact KPM data,
except for small $|M^{R}(\bar{\omega}=0,\bar{k}=0)|$ (see inset).
The Thomas-Fermi approximation clearly misses the phase transition
but is valid asymptotically and we note Ref. \cite{Pesin2015} suggesting
its systematic improvement for the 3d Weyl case, albeit for a different
disorder model. 

Second, we consider a 2d semimetal $H_{0}(\vec{k})=\frac{\hbar^{2}k^{2}}{m}\left[\sigma_{x}\cos(2\phi)+\sigma_{y}\sin(2\phi)\right]$
in polar coordinates. A similar Hamiltonian (with discrete rotation
symmetry) occurs for Bernal-stacked bilayer graphene \cite{McCann2013}.
The dispersion is parabolic, $E_{0}(k)=\hbar^{2}k^{2}/m$ and we have
$E_{\xi}=\hbar^{2}\xi^{-2}/m$ as the fundamental energy unit. The
SDWA (see supplement \cite{Note2} for details) yields good agreement to the KPM
data, see Fig. \ref{fig:res_3dWeyl_2dParabolic}(b), except in the
small $|M^{R}(\bar{\omega}=0,\bar{k}=0)|$ regime below $W\simeq2$. 

\emph{Derivation of main result.}\textemdash In the following, we
sketch the main ideas behind Eq. (\ref{eq:SD-Ward}). For a detailed
derivation we refer to the supplemental material \cite{Note2}. Let $\mathrm{ln}\,\mathcal{Z}_{V}[\bar{\eta},\eta]$
be the generating functional for connected Green functions for a fixed
disorder configuration $V$. The replica trick asserts that we can
obtain the disorder averaged Green functions from the generating functional
$\mathcal{Z}_{R}\left[\bar{\eta},\eta\right]=\left\langle \mathcal{Z}_{V}^{R}\left[\bar{\eta},\eta\right]\right\rangle _{dis}$
as $G_{12}=\underset{R\rightarrow0}{\mathrm{lim}}\frac{1}{R}\frac{\delta^{2}\mathcal{Z}_{R}\left[\bar{\eta},\eta\right]}{\delta\bar{\eta}_{1}\delta\eta_{2}}|_{\eta,\bar{\eta}=0}$,
where $\mathcal{Z}_{V}^{R}\left[\bar{\eta},\eta\right]$ contains
$R$ replicated fermion species $\psi^{\alpha}$, $\alpha=1,2,...,R$,
all coupling to the same disorder potential $V$ and the same source
$\bar{\eta}$ (analogous for $\bar{\psi}^{\alpha}$ and $\eta$).
We can now formally consider $\mathcal{Z}_{R}\left[\bar{\eta},\eta\right]\rightarrow\mathcal{Z}_{R}\left[\bar{\eta}^{\alpha},\eta^{\alpha},J\right]$
such that each fermion species couples to separate sources $\eta^{\alpha}$
and $\bar{\eta}^{\alpha}$ and introduce a source $J$ for the bosonic
field $V$. Now, the Green function from Eq. (\ref{eq:Sigma_def})
can be obtained as 
\begin{align}
G_{12} & =\underset{R\rightarrow0}{\mathrm{lim}}\frac{1}{R}\sum_{\alpha=1}^{R}G_{12}^{\alpha\alpha},\quad\!G_{12}^{\alpha\alpha}=\frac{\delta^{2}\mathcal{Z}_{R}\left[\bar{\eta}^{\alpha},\eta^{\alpha},J\right]}{\delta\bar{\eta}_{1}^{\alpha}\delta\eta_{2}^{\alpha}}\bigg|_{\bar{\eta},\eta,J=0},
\end{align}
and likewise for the self-energy $\Sigma_{12}$, which is obtained
as the second functional derivative of the generating functional of
irreducible vertex functions \cite{Kopietz2010}.

\begin{figure}
\noindent \begin{centering}
\includegraphics{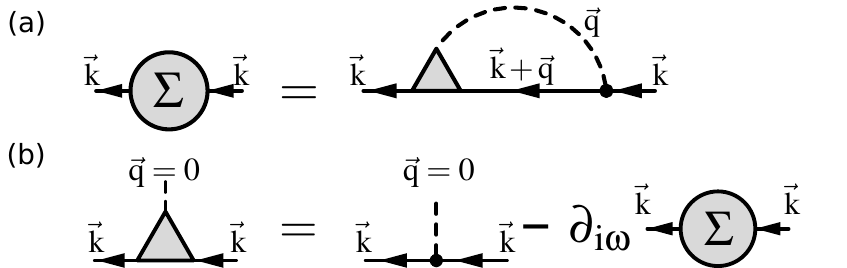}
\par\end{centering}
\caption{\label{fig:SD-Ward-diagrams}{\small{}(a) Schwinger-Dyson equation
for the }fermion{\small{}ic self-energy $\Sigma$ and (b) the Ward-identity
for the }Fermi-Bose{\small{} vertex (triangle). Together, they form
the basis of the proposed }SDW{\small{}A. We suppress frequency and
pseudo-spin indices.}}
\end{figure}

Next, we consider the Schwinger-Dyson equation for the self-energy,
shown diagrammatically in Fig. \ref{fig:SD-Ward-diagrams}(a). In
the diagram, we already anticipate the replica limit that eliminates
diagrammatic contributions with internal fermion loops \cite{Altland2006}.
In this way a Hartree-like diagram, still present for $\Sigma^{\alpha\alpha}$,
vanishes. Likewise, the bosonic self-energy, which contains internal
fermion bubbles, is eliminated in the replica limit, such that the
boson propagator $\mathcal{K}(\vec{q})$ is undressed (dashed line).
Since $\mathcal{K}(\vec{q})$ is related to elastic scattering, no
frequency is carried. The fermionic propagator in the loop on the
right hand side does involve the fermionic self-energy from the left
hand side. Finally, the triangle represents the Fermi-Bose vertex
that, besides its bare part, subsumes the effect of all higher order
diagrams with crossed impurity lines missing in the SCBA. 
In Fig. \ref{fig:SD-Ward-diagrams}(b) we depict a Ward-identity for
our theory $\mathcal{Z}_{R}$. It relates the Fermi-Bose vertex to
the Matsubara frequency derivative of the fermionic self-energy. This
relation follows from the invariance of the generating functional
$\mathcal{Z}_{R}\left[\bar{\eta}^{\alpha},\eta^{\alpha},J\right]$
under a temporal gauge transformation $\psi_{\sigma}^{\alpha}(\tau,\vec{r})=e^{+iA_{\alpha}(\tau)}\psi_{\sigma}^{\prime\alpha}(\tau,\vec{r})$
and $\bar{\psi}_{\sigma}^{\alpha}(\tau,\vec{r})=e^{-iA_{\alpha}(\tau)}\bar{\psi}_{\sigma}^{\prime\alpha}(\tau,\vec{r})$.
At an intermediate stage of the derivation, a bosonic Schwinger-Dyson
equation (not shown) is used. 

The idea is to eliminate the Fermi-Bose vertex in the Schwinger-Dyson
equation (a) using the Ward-identity (b). Crucially, the Ward-identity
requires vanishing bosonic momentum $\vec{q}=0$. Thus, in order to
use (b) in (a), we need to approximate the Fermi-Bose vertex with
its value at $\vec{q}\!=\!0$. Note that we keep $\vec{q}$ everywhere
else in the diagram. This yields Eq. (\ref{eq:SD-Ward}).

\begin{figure}
\noindent \begin{centering}
\includegraphics{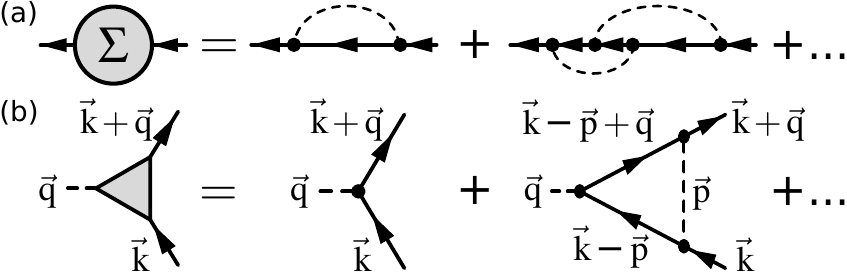}
\par\end{centering}
\caption{\label{fig:q0_justification}{\small{}(a) Self-consistent perturbation
theory for the disorder problem. The corresponding expansion of the
}Fermi-Bose{\small{} vertex is shown in (b). Up to second order, it
is used to argue for the validity of the $\vec{q}=0$ approximation
in the limit of strong disorder. All internal fermion lines are self-energy
dressed propagators. }}
\end{figure}

To motivate the above approximation, note that in the diagram of Fig.
\ref{fig:SD-Ward-diagrams}(a), we can restrict $|\vec{q}|=q\lesssim1/\xi$
due to the finite range of the bosonic propagator $\mathcal{K}(\vec{q})\sim e^{-q^{2}\xi^{2}/2}$.
We argue for the validity of the approximation on the basis of the
standard self-consistent expansion of the disorder self-energy \cite{BruusBook},
see Fig. \ref{fig:q0_justification}(a). Comparing to the Schwinger-Dyson
Eq. \ref{fig:SD-Ward-diagrams}(a), we obtain the expansion of the
Fermi-Bose vertex shown in Fig. \ref{fig:q0_justification}(b). Alternatively,
the expansion in Fig. \ref{fig:q0_justification}(b) can be obtained
from a Schwinger-Dyson equation for the Fermi-Bose vertex if the four-fermion
vertex is treated perturbatively. The bare contribution is a constant
and trivially $\vec{q}$-independent. The next contribution is a diagram
with an internal boson line. The value of the internal (dressed) fermion
propagators is dominant and nearly constant for momenta with magnitude
$\lesssim1/\gamma$, where $\gamma$ is the length-scale associated
to disorder broadening of the pole; a finite Matsubara frequency $\omega>0$
only increases $\gamma$. Our approximation is valid in the regime
$\gamma\lesssim\xi$, which means that the fermionic propagator with
momentum $\vec{k}-\vec{p}+\vec{q}$ does not change once $\vec{q}$
is set to zero. It is plausible that this argument holds for all higher
order diagrams although we cannot give a general proof. A priori,
the relation between disorder strength $W$ and $\gamma$ is not clear,
but the condition $\gamma\lesssim\xi$ can be checked from the result
of the SDW approach a posteriori. Note however, that keeping the full
momentum dependence of $G(\vec{k}+\vec{q})$ in the diagram of Fig.
\ref{fig:SD-Ward-diagrams}(a) is essential, setting $\vec{q}=0$
yields considerably worse results. 

\emph{Conclusion.}\textemdash We presented a non-perturbative approach
to calculate disorder averaged quasi-particle properties beyond SCBA.
The SDWA for the self-energy requires a sufficient broadening of the
quasiparticle pole to control the approximation involved. Systematic
improvement is possible using a higher level truncation of the Schwinger-Dyson
equations. This extended set may then be closed by Ward-identities.
This should also allow for the calculation of conductivities. In contrast
to the numerically expensive KPM, the analytical formulation of the
SDW makes this method amenable for integration in other, possibly
interacting, theories. For future work, one could try to the apply
the SDWA to other types of disorder with non-trivial Pauli matrix
structure \cite{Ostrovsky2006,Sbierski2016b} or relax the particle-hole
symmetry and isotropy assumption on the dispersion to study disordered
tilted or anisotropic cones \cite{Trescher2015,Trescher2016}. The
SDWA could also be useful for semimetals with a co-dimension of their
Fermi-surface smaller than $d$, for example nodal-line semimetals
in 3d \cite{Burkov2011a}. 

\emph{Acknowledgments.}\textemdash We thank Jörg Behrmann, Piet Brouwer,
Christoph Karrasch, Georg Schwiete and Sergey Syzranov for useful
discussions. Numerical computations were done on the HPC cluster of
Fachbereich Physik at FU Berlin. Financial support was granted by
the Deutsche Forschungsgemeinschaft through the Emmy Noether-program
(KA 3360/2-1) and the CRC/Transregio 183 (Projects A02 and A03).

\onecolumngrid
\noindent \begin{center}
\textbf{\large{}\uline{Supplemental material}}{\large\par}
\par\end{center}

\noindent \begin{center}
\textbf{\large{}for ``Strong disorder in nodal semimetals: Schwinger-Dyson\textendash Ward
approach''}{\large\par}
\par\end{center}

\section*{{\large{}Weak disorder in 2d Dirac node}}

\begin{figure}[H]
\noindent \begin{centering}
\includegraphics{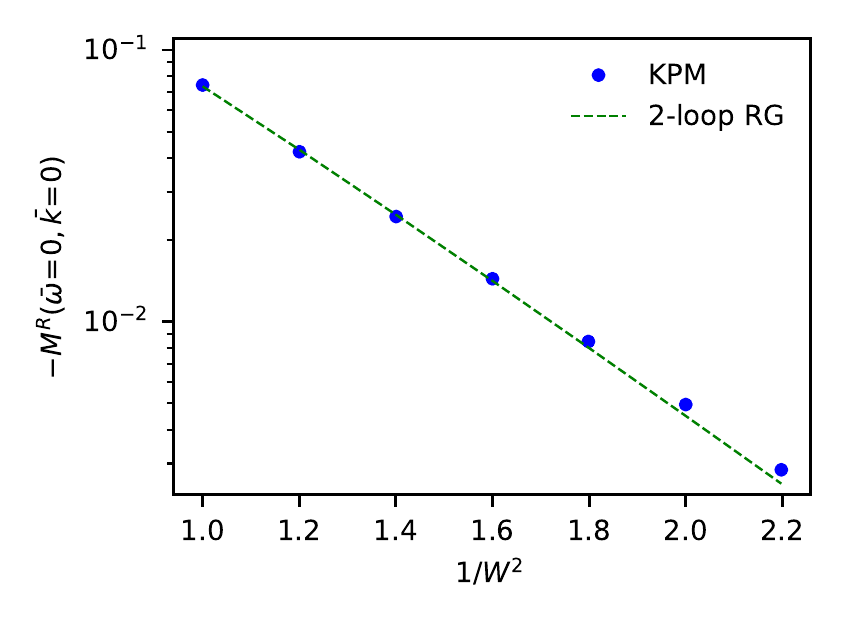}
\par\end{centering}
\caption{{\small{}\label{fig:res_2dDirac_GWN}Imaginary part of the disorder
induced retarded self-energy for a 2d Dirac node with discrete disorder
as a function of disorder strength $W$. The exact KPM data is shown
as blue dots, the RG prediction $M^{R}\sim W^{-1}\exp(-\pi/W^{2})$
with an appropriate prefactor fitted is shown as a green dashed line. }}
\end{figure}

\noindent We now consider weak disorder in a 2d Dirac node. In Fig.
\ref{fig:res_2dDirac_GWN} we prove by comparison to exact KPM data
(blue dots) that the scaling $M^{R}(\bar{\omega}=0,\bar{k}=0)\sim W^{-1}\exp(-\pi/W^{2})$
inferred from the 2-loop momentum shell RG flow equation \cite{Schuessler2009}
is correct (green dashed line). This result is obtained from the scale
where the RG-flow crosses over to strong disorder. To the best of
our knowledge, this scaling has never been checked numerically. Note
that the RG uses a white-noise disorder correlator which cannot be
implemented numerically and is responsible for the $\sim$ sign above.
To obtain converged values of small $M^{R}$ over two orders of magnitude,
we chose a discrete disorder model where the ``correlation length''
$\xi$ equals the real-space lattice constant $a$, such that the
disorder correlator $\xi=a$ is not smooth on the lattice scale. Thus,
the SDWA formulated for the field-theory limit $\xi\gg a$ is not
directly applicable. The potential at each site of the real-space
lattice is uniformly distributed, $V(\vec{r})\in[-\sqrt{3}W,\sqrt{3}W]$.
This yields $\sum_{\vec{r}}\left\langle V(\vec{r})V(\vec{0})\right\rangle _{dis}=\frac{1}{2\sqrt{3}W}\int_{-\sqrt{3}W}^{+\sqrt{3}W}dU\,U^{2}=W^{2}$.
We use $P=2^{12}$ lattice points in both linear directions, such
that $a=\frac{2\pi}{P\Delta k}$ and $60000$ moments for convergence
of the KPM. We checked that the disorder induced energy scale $M^{R}(\bar{\omega}=0,\bar{k}=0)\hbar v/a$
is always larger than the finite-size energy scale $E_{fs}=\hbar v\Delta k$.
\noindent 

\section*{{\large{}Thomas-Fermi approximation for strong disorder}}

\noindent The Thomas-Fermi approximation \cite{Pesin2015,Shklovskii1984}
amounts to approximate the disorder potential as a homogeneous effective
chemical potential term. Then, the disorder averaged Green function
is approximated as
\begin{equation}
G^{R}(\omega,\vec{k})=\int_{-\infty}^{\infty}dU\,P_{1}(U)\frac{1}{\omega+i\eta-H_{0}(\vec{k})-U},\label{eq:ThomasFermi}
\end{equation}
where $P_{1}(U)$ is the one-point distribution function, i.e. the
probability that the local potential $V(\vec{r}_{1})$ has the value
$U$. This probability can be obtained as the expectation value $P_{1}(U)=\left\langle \delta\left(V(\vec{r}_{1})-U\right)\right\rangle _{dis}$
which can be calculated by representing the $\delta$-function as
an integral over an exponential and subsequently employing the rules
for functional integration over $V$. The result is
\begin{eqnarray}
P_{1}(U) & = & \left\langle \frac{1}{2\pi}\int_{-\infty}^{+\infty}dx\,e^{i\left(V(\vec{r}_{1})-U\right)x}\right\rangle _{dis}=\frac{1}{2\pi}\int_{-\infty}^{+\infty}dx\,e^{-\frac{1}{2}x^{2}\left\langle V^{2}(\vec{r}_{1})\right\rangle _{dis}-iUx}=\frac{1}{\sqrt{2\pi\mathcal{K}(\vec{0})}}e^{-\frac{U^{2}}{2\mathcal{K}(\vec{0})}},
\end{eqnarray}
where $\mathcal{K}(\vec{0})=\left\langle V^{2}(\vec{r}_{1})\right\rangle _{dis}=(2\pi)^{-d/2}W^{2}E_{\xi}^{2}$.

We evaluate Eq. (\ref{eq:ThomasFermi}) for the 2d Dirac case in the
helicity basis at $\omega=0$ and find
\begin{eqnarray}
G_{\lambda,\lambda^{\prime}}^{R}(\bar{\omega}=0,\bar{k})E_{\xi} & = & -\delta_{\lambda,\lambda^{\prime}}\frac{\pi}{W}\left(\text{erfi}\left(\frac{\lambda\bar{k}}{\sqrt{\pi}W}\right)+i\right)e^{-(\bar{k})^{2}/(\pi W^{2})}.
\end{eqnarray}
Here, $\text{erfi}(z)$ is the imaginary error function defined as
$\text{erfi}(z)=\text{erf}(iz)/i$ \cite{Gradshteyn2007}. With the
ansatz 
\begin{equation}
G_{\lambda,\lambda^{\prime}}^{R}(\bar{\omega}=0,\bar{k})E_{\xi}=\frac{\delta_{\lambda,\lambda^{\prime}}}{-\lambda\left[\bar{k}+m^{R}(\bar{\omega}=0,\bar{k})\right]-iM^{R}(\bar{\omega}=0,\bar{k})},
\end{equation}
we obtain
\begin{eqnarray}
M^{R}(\bar{\omega}=0,\bar{k}) & = & -\frac{W}{\pi}\frac{e^{\bar{k}^{2}/(\pi w^{2})}}{\text{erfi}^{2}\left(\frac{\bar{k}}{\sqrt{\pi}W}\right)+1}=-\frac{W}{\pi}+\frac{(4-\pi)\bar{k}^{2}}{W\pi^{3}}+\mathcal{O}(\bar{k}^{3}),\\
m^{R}(\bar{\omega}=0,\bar{k}) & = & -\bar{k}+\frac{W}{\pi}\frac{e^{\bar{k}^{2}/(\pi w^{2})}\text{erfi}\left(\frac{\bar{k}}{\sqrt{\pi}W}\right)}{\text{erfi}^{2}\left(\frac{\bar{k}}{\sqrt{\pi}W}\right)+1}=-\underset{\simeq0.8}{\underbrace{(1-2/\pi^{2})}}\bar{k}+\mathcal{O}(\bar{k}^{2}).
\end{eqnarray}
\\

\section*{{\large{}Details on the SDWA for 3d Weyl and parabolic 2d semimetal}}

\noindent For the 3d Weyl node with $H_{0}(\vec{k})=\hbar v(\sigma_{x}k_{x}+\sigma_{y}k_{y}+\sigma_{z}k_{z})$
we consider the self-energy ansatz
\begin{align}
\Sigma(\vec{k},i\omega)/E_{\xi}= & m(\bar{\omega},\bar{k})\left[\sin\bar{\theta}\left(\sigma_{x}\cos\bar{\phi}+\sigma_{y}\sin\bar{\phi}\right)+\sigma_{z}\cos\bar{\theta}\right]+iM(\bar{\omega},\bar{k}).\label{eq:Sigma_3dWeyl}
\end{align}
Upon insertion into Eq. (3) we find that Eqs. (5)
and (6) remain valid, but with the replacements $\{J_{0},J_{1}\}$
$\rightarrow$ $\{J_{0}^{w},J_{1}^{w}\}$ where
\begin{align}
\left\{ \begin{array}{c}
\!J_{0}^{w}(\bar{\omega},\bar{k})\!\\
\!J_{1}^{w}(\bar{\omega},\bar{k})\!
\end{array}\right\} = & \!\int_{0}^{\infty}\!\!d\bar{q}\left\{ \begin{array}{c}
\bar{k}\bar{q}\sinh(\bar{k}\bar{q})\tilde{M}(\bar{\omega},\bar{q})\\
\!\!\left[\bar{k}\bar{q}\cosh(\bar{k}\bar{q})-\sinh(\bar{k}\bar{q})\right]\tilde{m}(\bar{\omega},\bar{q})\!\!
\end{array}\right\} \frac{e^{-(\bar{q}^{2}+\bar{k}^{2})/2}/(2\pi^{2}\bar{k}^{2})}{\tilde{m}^{2}(\bar{\omega},\bar{q})+\tilde{M}^{2}(\bar{\omega},\bar{q})}.\label{eq:J_3dWeyl}
\end{align}
In addition, the initial conditions are modified to $M(\bar{\omega}_{max},\bar{k})=-W^{2}(2\pi)^{-3/2}/\bar{\omega}_{max}$
and $m(\bar{\omega}_{max},\bar{k})=0$.

For the 2d parabolic semimetal, described by the clean Hamiltonian
$H_{0}(\vec{k})=\frac{\hbar^{2}k^{2}}{m}\left[\sigma_{x}\cos(2\phi)+\sigma_{y}\sin(2\phi)\right]$
we use the self-energy ansatz 
\begin{equation}
\Sigma(\vec{k},i\omega)/E_{\xi}=m(\bar{\omega},\bar{k})\left[\sigma_{x}\cos(2\bar{\phi})+\sigma_{y}\sin(2\bar{\phi})\right]+iM(\bar{\omega},\bar{k}).\label{eq:Sigma_2dParabolic}
\end{equation}
We arrive at Eqs. (5) and (6) with the redefinition
$\tilde{m}(\bar{\omega},\bar{k})=\bar{k}^{2}+m(\bar{\omega},\bar{k})$
and the replacements $\{J_{0},J_{1}\}$ $\rightarrow$ $\{J_{0}^{p},J_{1}^{p}\}$,
where
\begin{equation}
\left\{ \begin{array}{c}
\!J_{0}^{p}(\bar{\omega},\bar{k})\!\\
\!J_{1}^{p}(\bar{\omega},\bar{k})\!
\end{array}\right\} \!=\!\int_{0}^{\infty}\!\!d\bar{q}\left\{ \begin{array}{c}
\!\!\tilde{M}(\bar{\omega},\bar{q})I_{0}(\bar{k}\bar{q})\!\!\\
\!\!\tilde{m}(\bar{\omega},\bar{q})I_{2}(\bar{k}\bar{q})\!\!
\end{array}\right\} \frac{\bar{q}e^{-(\bar{q}^{2}+\bar{k}^{2})/2}/(2\pi)}{\tilde{m}^{2}(\bar{\omega},\bar{q})+\tilde{M}^{2}(\bar{\omega},\bar{q})}.\label{eq:J_2dParabolic}
\end{equation}
The initial conditions are the same as in the 2d Dirac case, $M(\bar{\omega}_{max},\bar{k})=-\frac{W^{2}}{2\pi\bar{\omega}_{max}}$
and $m(\bar{\omega}_{max},\bar{k})=0$.\\

\section*{{\large{}Detailed derivation of the Schwinger-Dyson-Ward approximation}}

\noindent Within the following derivation, we mostly stick to the
conventions and definitions of Ref. \cite{Kopietz2010}.

\subsection{Preliminaries}

\label{sec:Preliminaries} For a given disorder realization $V(\vec{r})$,
the generating functional for fermionic imaginary-time Green functions
is given by 
\begin{equation}
\mathcal{Z}_{V}\left[\bar{\eta},\eta\right]=\int\mathcal{D}\bar{\psi}\,\mathcal{D}\psi\,e^{-S_{V}\left[\bar{\psi},\psi\right]+(\bar{\eta},\psi)+(\bar{\psi},\eta)},\label{eq:ZV}
\end{equation}
where $S_{V}\left[\bar{\psi},\psi\right]$ is the Euclidean action
of the system in the presence of the disorder field $V$ 
\begin{equation}
S_{V}\left[\bar{\psi},\psi\right]=S_{0}\left[\bar{\psi},\psi\right]+\sum_{\sigma}\int_{\vec{r},\tau}\bar{\psi}_{\sigma}\left(\vec{r},\tau\right)V\left(\vec{r}\right)\psi_{\sigma}\left(\vec{r},\tau\right).\label{eq:DisorderAction}
\end{equation}
The index $\sigma$ plays the role of a pseudo-spin, which is necessary
to describe a two band model. Its clean part, $S_{0}\left[\bar{\psi},\psi\right]$,
derives from the Hamiltonian $H_{0}(\vec{k}=-i\hbar\vec{\nabla})$
as follows \cite{Negele1988,Altland2006}
\begin{equation}
S_{0}\left[\bar{\psi},\psi\right]=\sum_{\sigma\sigma'}\int_{\vec{r},\tau}\bar{\psi}_{\sigma}\left(\vec{r},\tau\right)\left[\partial_{\tau}+H_{0}(-i\hbar\vec{\nabla})\right]_{\sigma\sigma^{\prime}}\psi_{\sigma^{\prime}}\left(\vec{r},\tau\right).
\end{equation}
For the source terms, involving the Grassmann-valued fields $\eta$
and $\bar{\eta}$, we used the compact scalar product notation $(\bar{\psi},\eta)\equiv\sum_{\sigma}\int_{\vec{r},\tau}\bar{\psi}_{\sigma}\left(\vec{r},\tau\right)\eta_{\sigma}\left(\vec{r},\tau\right)$.
The $n$-point Green functions at a fixed disorder field configuration
can then be obtained as an $n$-fold functional derivative of $\mathcal{Z}_{V}$
with respect to the sources. For the two-point Green function, for
example, we have
\begin{equation}
G_{V,\sigma\sigma^{\prime}}\left(\vec{r},\tau;\vec{r}^{\prime},\tau^{\prime}\right)=-\left\langle \psi_{\sigma}(\vec{r},\tau)\bar{\psi}_{\sigma^{\prime}}(\vec{r}^{\prime},\tau^{\prime})\right\rangle =\frac{1}{\mathcal{Z}_{V}[0,0]}\frac{\delta^{2}\mathcal{Z}_{V}\left[\bar{\eta},\eta\right]}{\delta\bar{\eta}_{\sigma}(\vec{r},\tau)\,\delta\eta_{\sigma^{\prime}}(\vec{r}^{\prime},\tau^{\prime})}\bigg|_{\eta=\bar{\eta}=0}.\label{eq:FixedDisorderTwoPointFunction}
\end{equation}
Note the appearance of the $V$-dependent normalization $\mathcal{Z}_{V}[0,0]$.
The connected $n$-point Green functions at fixed disorder configuration
are defined as the $n$-fold derivative of the connected functional
$\mathcal{G}_{V}\left[\bar{\eta},\eta\right]=\ln\mathcal{Z}_{V}\left[\bar{\eta},\eta\right]$,
just as usual. Since our theory is non-interacting, the only non-vanishing
connected correlator is the two-point function~(\ref{eq:FixedDisorderTwoPointFunction}).

\noindent Since we are not interested in one particular disorder realization,
but in a statistical ensemble of disorder potentials, we have to perform
an ensemble average. To this end, one has to specify the statistical
properties of the ensemble, which are summarized in a probability
distribution $P[V]$. Here, we choose the Gaussian probability distribution
\begin{equation}
P\left[V\right]=\mathcal{N}\exp\left(-\frac{1}{2}\int_{\vec{r},\vec{r}^{\prime}}V(\vec{r})\mathcal{K}^{-1}(\vec{r}-\vec{r}^{\prime})V(\vec{r}^{\prime})\right),
\end{equation}
where $\mathcal{K}^{-1}\left(\vec{r}-\vec{r}^{\prime}\right)$ is
the distributional inverse of the fundamental disorder correlator
$\left\langle V(\vec{r})V(\vec{r}^{\prime})\right\rangle _{dis}=\mathcal{K}\left(\vec{r}-\vec{r}^{\prime}\right)$,
and $\mathcal{N}$ is a normalization constant. The disorder average
of a general quantity $Q[V]$ is then defined as $\left\langle Q\right\rangle _{dis}=\int\mathcal{D}VQ[V]P[V]$.
To obtain disorder averaged correlation functions one would have to
either average each $n$-point function individually, or average the
normalized generating functional~(\ref{eq:ZV}), that is $\left\langle \mathcal{Z}_{V}[\bar{\eta},\eta]/\mathcal{Z}_{V}[0,0]\right\rangle _{dis}$,
which would serve as the generating functional of disorder averaged
Green functions. However, due to the $V$-dependent normalization
$\mathcal{Z}_{V}[0,0]$ in the denominator, it is not possible to
naively perform the disorder average.

To circumvent this problem and get rid of the denominator there are
three possibilities \cite{Altland2006}: (1) work in real time using
the Keldysh technique; (2) rewrite the denominator as a bosonic Gaussian
integral, a technique known as supersymmetry method; or (3) ``replicate
the system'' and perform an analytical continuation to zero replicas
at the end. Here, we choose the latter option. The trick is to rewrite
the connected functional $\mathcal{G}_{V}\left[\bar{\eta},\eta\right]$
as follows 
\begin{equation}
\mathcal{G}_{V}\left[\bar{\eta},\eta\right]=\ln\mathcal{Z}_{V}\left[\bar{\eta},\eta\right]=\underset{R\rightarrow0}{\mathrm{lim}}\frac{1}{R}\left(e^{R\ln\mathcal{Z}_{V}\left[\bar{\eta},\eta\right]}-1\right)=\underset{R\rightarrow0}{\mathrm{lim}}\frac{1}{R}\left(\mathcal{Z}_{V}^{R}\left[\bar{\eta},\eta\right]-1\right),\label{eq:replica}
\end{equation}
which allows us to perform the disorder average. The disorder averaged
replicated partition function then reads 
\begin{align}
\mathcal{Z}_{R}\left[\bar{\eta},\eta\right] & \equiv\left\langle \mathcal{Z}_{V}^{R}\left[\bar{\eta},\eta\right]\right\rangle _{dis}\nonumber \\
 & =\int\mathcal{D}\bar{\psi}^{\alpha}\,\mathcal{D}\psi^{\alpha}\,\mathcal{D}V\exp\left(-S_{R}[\bar{\psi}^{\alpha},\psi^{\alpha},V]+\sum_{\alpha=1}^{R}(\bar{\eta},\psi^{\alpha})+\sum_{\alpha=1}^{R}(\bar{\psi}^{\alpha},\eta)\right),\label{eq:ZR}
\end{align}
with the replicated action $S_{R}[\bar{\psi}^{\alpha},\psi^{\alpha},V]\equiv\sum_{\alpha=1}^{R}S_{V}[\bar{\psi}^{\alpha},\psi^{\alpha}]+\frac{1}{2}\int_{\vec{r},\vec{r}'}V(\vec{r})\mathcal{K}^{-1}(\vec{r}-\vec{r}')V(\vec{r}')$.
Note that there is only a single source for all replicas and only
a single disorder potential coupling identically to the replica bilinears
in $S_{V}[\bar{\psi}^{\alpha},\psi^{\alpha}]$. In the standard treatment
one would integrate out the disorder field to obtain a quartic pseudo-interaction
term for the fermions \cite{Altland2006}, but here we take another
path. Instead of performing the bosonic Gaussian integral, we consider
a generalization of Eq.~(\ref{eq:ZR}), where a bosonic source $J$
coupling to $V$ is introduced and where the fermionic sources now
carry a replica index as well 
\begin{equation}
\mathcal{Z}_{R}\left[\bar{\eta},\eta\right]\rightarrow\mathcal{Z}_{R}\left[\bar{\eta}^{\alpha},\eta^{\alpha},J\right]=\int\mathcal{D}\bar{\psi}^{\alpha}\,\mathcal{D}\psi^{\alpha}\,\mathcal{D}V\exp\left(-S_{R}\left[\bar{\psi}^{\alpha},\psi^{\alpha},V\right]+\sum_{\alpha=1}^{R}(\bar{\eta}^{\alpha},\psi^{\alpha})+\sum_{\alpha=1}^{R}(\bar{\psi}^{\alpha},\eta^{\alpha})+(J,V)\right).\label{eq:GeneralizedZR}
\end{equation}
Introducing the super-field vector $\boldsymbol{\Phi}=(\psi^{\alpha},\bar{\psi}^{\alpha},V)$,
the super-source vector $\mathbf{J}=\left(\bar{\eta}^{\alpha},-\eta^{\alpha},J\right)$,
and the scalar product $(\mathbf{J},\boldsymbol{\Phi})=\sum_{\alpha=1}^{R}(\bar{\eta}^{\alpha},\psi^{\alpha})+\sum_{\alpha=1}^{R}(\bar{\psi}^{\alpha},\eta^{\alpha})+(J,V)$,
we can write this new functional in the compact form $\mathcal{Z}_{R}\left[\mathbf{J}\right]\equiv\int\mathcal{D}\boldsymbol{\Phi}\exp\left(-S_{R}\left[\boldsymbol{\Phi}\right]+(\mathbf{J},\boldsymbol{\Phi})\right)$.
Putting everything together we find the disorder averaged connected
Green function 
\begin{equation}
\left\langle G_{V,12}\right\rangle _{dis}=\left\langle \frac{\delta^{2}\ln\mathcal{Z}_{V}\left[\bar{\eta},\eta\right]}{\delta\bar{\eta}_{1}\delta\eta_{2}}\bigg|_{\bar{\eta}=\eta=0}\right\rangle _{dis}\overset{\eqref{eq:replica}-\eqref{eq:GeneralizedZR}}{=}\underset{R\rightarrow0}{\mathrm{lim}}\frac{1}{R}\sum_{\alpha=1}^{R}\frac{\delta^{2}\mathcal{Z}_{R}\left[\mathbf{J}\right]}{\delta\bar{\eta}_{1}^{\alpha}\delta\eta_{2}^{\alpha}}\bigg|_{\mathbf{J}=0}\equiv\underset{R\rightarrow0}{\mathrm{lim}}\frac{1}{R}\sum_{\alpha=1}^{R}G_{12}^{\alpha\alpha}.\label{eq:Physical vs Replica}
\end{equation}
Here, the numerical indices 1 and 2 are a compact notation, which
include space, imaginary-time and pseudo-spin indices. In the following
we consider a finite number of replicas $R$ \textendash{} the replica
limit will only be taken at the end of the calculation \textendash{}
and compute $G_{12}^{\alpha\alpha}=\frac{\delta^{2}\mathcal{Z}_{R}\left[\mathbf{J}\right]}{\delta\bar{\eta}_{1}^{\alpha}\delta\eta_{2}^{\alpha}}\big|_{\mathbf{J}=0}=-\left\langle \psi_{1}^{\alpha}\bar{\psi}_{2}^{\alpha}\right\rangle _{S_{R}}$.
The subscript at the average is just a reminder that it has to be
performed with respect to the replicated action $S_{R}$ in Eq.~(\ref{eq:GeneralizedZR}).

\subsection{Schwinger-Dyson equations}

\label{sec:SchwingerDysonEquations} As is well-known, in classical
field theory the equations of motions follow from a least action principle.
Its generalization to quantum field theories and the corresponding
quantum equations of motion follow from a functional analog of the
fundamental theorem of calculus, stating that the functional integral
over a total derivative vanishes \cite{Peskin1995,Kopietz2010}
\begin{align}
0=\int\mathcal{D}\boldsymbol{\Phi}\frac{\delta}{\delta\boldsymbol{\Phi}_{i}}e^{-S_{R}\left[\boldsymbol{\Phi}\right]+(\mathbf{J},\boldsymbol{\Phi})}=\int\mathcal{D}\boldsymbol{\Phi}\left(-\frac{\delta S_{R}}{\delta\boldsymbol{\Phi}_{i}}+\boldsymbol{\zeta}_{i}\boldsymbol{J}_{i}\right)e^{-S_{R}\left[\boldsymbol{\Phi}\right]+(\mathbf{J},\boldsymbol{\Phi})}\equiv\left\langle -\frac{\delta S_{R}}{\delta\boldsymbol{\Phi}_{i}}+\boldsymbol{\zeta}_{i}\boldsymbol{J}_{i}\right\rangle _{\mathbf{J}}.\label{eq:SchwingerDyson}
\end{align}
Here, the index $i$ encompasses space, imaginary time, the discrete
pseudo-spin index as well as the super-field component, see our definition
above. Furthermore, $\boldsymbol{\zeta}_{i}$ is a statistical factor,
which is $-1$ for a fermionic source and $+1$ for a bosonic one,
and the subscript $\mathbf{J}$ at the functional average indicates
that it has to be performed in the presence of the source fields.
We can rewrite these expectation values as functional differential
equations by replacing the $\boldsymbol{\Phi}$-dependence of $\frac{\delta S_{R}}{\delta\boldsymbol{\Phi}_{i}}$
with their corresponding source-derivatives 
\begin{equation}
\left(-\frac{\delta S_{R}}{\delta\boldsymbol{\Phi}_{i}}\left[\frac{\delta}{\delta\mathbf{J}}\right]+\boldsymbol{\zeta}_{i}\boldsymbol{J}_{i}\right)\mathcal{Z}_{R}\left[\mathbf{J}\right]=0.\label{eq:master-SD}
\end{equation}
This set of equations is known as Schwinger-Dyson equations and they
serve as master equations, which can be functionally differentiated
to obtain an infinite hierarchy of coupled integral equations for
the one-particle irreducible vertex functions.

\noindent To obtain such equations one has to switch to the connected
functional $\mathcal{G}_{R}[\mathbf{J}]=\ln\mathcal{Z}_{R}[\mathbf{J}]$,
and perform a Legendre transformation to the effective action $\mathcal{L}_{R}[\mathbf{\Phi}]=-\mathcal{G}_{R}[\mathbf{J}]+(\mathbf{J},\mathbf{\Phi})$.
Here, the super-field vector $\mathbf{\Phi}$ is the quantum expectation
value $\mathbf{\Phi}=\frac{\delta\mathcal{G}_{R}[\mathbf{J}]}{\delta\mathbf{J}}$
in the presence of the source fields $\mathbf{J}$. It shall not be
confused with the integration variables in Eqs.~(\ref{eq:GeneralizedZR})
and~(\ref{eq:SchwingerDyson}). Following Ref. \cite{Kopietz2010},
we write $\mathcal{L}_{R}[\mathbf{\Phi}]=S_{0}[\mathbf{\Phi}]+\Gamma_{R}[\mathbf{\Phi}]$,
where $S_{0}[\mathbf{\Phi}]\equiv\sum_{\alpha=1}^{R}S_{V=0}[\bar{\psi}^{\alpha},\psi^{\alpha}]+\frac{1}{2}\int_{\vec{r},\vec{r}'}V(\vec{r})\mathcal{K}^{-1}(\vec{r}-\vec{r}')V(\vec{r}')$
is the bare quadratic action and $\Gamma_{R}[\mathbf{\Phi}]$ is the
generating functional of one-particle irreducible vertex functions,
or vertex functional for short. As a result, we find the Schwinger-Dyson
equations in the form
\begin{align}
\frac{\delta\Gamma_{R}[\mathbf{\Phi}]}{\delta\bar{\psi}_{\sigma}^{\alpha}(\vec{r},\tau)} & =V(\vec{r})\psi_{\sigma}^{\alpha}(\vec{r},\tau)+\frac{\delta^{2}\mathcal{G}_{R}[\mathbf{J}]}{\delta J(\vec{r})\delta\bar{\eta}_{\sigma}^{\alpha}(\vec{r},\tau)},\label{eq:VertexSchwingerDysonA}\\
\frac{\delta\Gamma_{R}[\mathbf{\Phi}]}{\delta\psi_{\sigma}^{\alpha}(\vec{r},\tau)} & =-\bar{\psi}_{\sigma}^{\alpha}(\vec{r},\tau)V(\vec{r})+\frac{\delta^{2}\mathcal{G}_{R}[\mathbf{J}]}{\delta\eta_{\sigma}^{\alpha}(\vec{r},\tau)\delta J(\vec{r})},\label{eq:VertexSchwingerDysonB}\\
\frac{\delta\Gamma_{R}[\mathbf{\Phi}]}{\delta V(\vec{r})} & =\sum_{\alpha=1}^{R}\sum_{\sigma}\int_{\tau}\bar{\psi}_{\sigma}^{\alpha}(\vec{r},\tau)\psi_{\sigma}^{\alpha}(\vec{r},\tau)-\sum_{\alpha=1}^{R}\sum_{\sigma}\int_{\tau}\frac{\delta^{2}\mathcal{G}_{R}[\mathbf{J}]}{\delta\eta_{\sigma}^{\alpha}(\vec{r},\tau)\delta\bar{\eta}_{\sigma}^{\alpha}(\vec{r},\tau)}.\label{eq:VertexSchwingerDysonC}
\end{align}
On the right hand side, the second functional derivatives of $\mathcal{G}_{R}$
still have to be replaced by second functional derivatives of $\Gamma_{R}$,
using the inversion relation between the Hesse matrices of $\mathcal{G}_{R}$
and $\mathcal{L}_{R}$, see Refs. \cite{Negele1988,Kopietz2010}.
This substitution eliminates the remaining source field dependence,
but it leads to rather complex expressions. For this reason we leave
the above equations in this compact mixed form. To obtain the infinite
hierarchy of integral equations for the one-particle irreducible vertex
functions as advertised above, one has to expand the vertex functional
$\Gamma_{R}$ in a Taylor series in terms of fields, insert the expression
on the left and right hand sides and compare coefficients. Alternatively
one may simply apply a corresponding amount of field derivatives $\frac{\delta}{\delta\mathbf{\Phi}}$
to the above set of equations and set the sources $\mathbf{J}$ to
zero afterwards. When the sources $\mathbf{J}$ are set to zero, the
fields in $\Gamma_{R}$ are set to their possibly finite expectation
value $\mathbf{\Phi_{c}}=\mathbf{\Phi}|_{\mathbf{J}=0}=\frac{\delta\mathcal{G}[\mathbf{J}]}{\delta\mathbf{J}}\big|_{\mathbf{J}=0}$
\cite{Kopietz2010}. (In the present case only the bosonic field may
develop a finite expectation value.) In the Taylor expansion of $\Gamma_{R}$
one should account for that fact by expanding around $\mathbf{\Phi}=\mathbf{\Phi_{c}}$,
instead of $\mathbf{\Phi}=0$, such that the vertex functions are
defined as field-derivatives of $\Gamma_{R}$ evaluated at $\mathbf{\Phi}=\mathbf{\Phi_{c}}$.

The Schwinger-Dyson equation for the fermionic self-energy, which
is defined by $\Sigma_{12}=-\delta^{2}\Gamma_{R}/\delta\bar{\psi}_{1}\delta\psi_{2}|_{\mathbf{\Phi}=\mathbf{\Phi_{c}}}$,
may be obtained from Eq.~(\ref{eq:VertexSchwingerDysonA}) after
applying the derivative $\frac{\delta}{\delta\psi}$. A short calculation
yields the following equation in Fourier space 
\begin{align}
\Sigma_{\sigma_{1}\sigma_{2}}^{\alpha\alpha}\left(i\omega,\vec{k}\right) & =\delta_{\sigma_{1},\sigma_{2}}\mathcal{K}(0)\sum_{\beta=1}^{R}\sum_{\sigma}\int_{\vec{k}',\omega'}G_{\sigma\sigma}^{\beta\beta}(i\omega',\vec{k}')\nonumber \\
 & -\sum_{\sigma}\int_{\vec{q}}F\left(\vec{q}\right)\frac{\delta^{3}\Gamma}{\delta\bar{\psi}_{\sigma_{1}}^{\alpha}(i\omega,\vec{k})\delta\psi_{\sigma}^{\alpha}(i\omega,\vec{k}+\vec{q})\delta V(-\vec{q})}\bigg|_{\mathbf{\Phi}=\mathbf{\Phi}_{c}}G_{\sigma\sigma_{2}}^{\alpha\alpha}(i\omega,\vec{k}+\vec{q}),\label{eq:SD_Sigma}
\end{align}
with $\int_{\vec{q}}=\int\frac{d^{d}q}{(2\pi)^{d}}$. The term in
the first line, involving a closed fermion loop and the bare disorder
propagator at vanishing momentum, is the Hartree contribution to the
self-energy. It represents the influence of a finite expectation value
of $V$ on the fermions, and has been obtained by employing Eq.~(\ref{eq:VertexSchwingerDysonC})
at $\mathbf{J}=0$. The term in the second line represents the Fock
exchange self-energy, where the third derivative of $\Gamma_{R}$
is the full Fermi-Bose three vertex. Furthermore, $G_{\sigma_{1}\sigma_{2}}^{\alpha\alpha}(i\omega,\vec{k})$
and $F\left(\vec{q}\right)$ are the full fermionic and bosonic propagators,
respectively. The former involves the fermionic self-energy, which
makes Eq.~(\ref{eq:SD_Sigma}) a self-consistency equation, while
the latter involves the bosonic self-energy \textendash{} polarization
bubbles, for which there exists a separate equation, that derives
from Eq.~(\ref{eq:VertexSchwingerDysonC}) after applying $\frac{\delta}{\delta V}$.
We emphasize that the closed fermion loops in the Hartree term and
the polarization bubbles are finite prior to taking the replica limit.
They only vanish in the replica limit, which we will discuss at the
end, see Sec.~\ref{sec:ReplicaLimit}. Anticipating the replica limit,
Eq. (\ref{eq:SD_Sigma}) is depicted in Fig. 3(a).

\subsection{Ward identity}

\label{sec:WardIdentity} According to Noether's theorem a continuous
symmetry in a classical field theory leads to conservation laws. In
a quantum field theory such symmetries lead to Ward identities, which
connect various vertex functions to one another \cite{Kopietz2010}.
In the present case the fermions obey a global $U(1)^{\otimes R}$
symmetry, which formally expresses the fact that the particle number
for each replica is conserved. To obtain a relation between different
correlation functions we have to consider a local $U(1)^{\otimes R}$
symmetry transformation 
\begin{equation}
\psi_{\sigma}^{\alpha}(\tau,\vec{r})=e^{+iA^{\alpha}(\tau)}\psi_{\sigma}^{\prime\alpha}(\tau,\vec{r}),\quad\bar{\psi}_{\sigma}^{\alpha}(\tau,\vec{r})=\bar{\psi}_{\sigma}^{\prime\alpha}(\tau,\vec{r})e^{-iA^{\alpha}(\tau)}.
\end{equation}
(Here, we took the phase field $A^{\alpha}$ to be local in imaginary
time only. Spatial locality is not relevant, but could be incorporated
without problems.) This transformation leads to an additional term
in the action, 
\begin{equation}
S_{R}\left[\bar{\psi}^{\alpha},\psi^{\alpha},V\right]=S_{R}\left[\bar{\psi}^{\prime\alpha},\psi^{\prime\alpha},V\right]+\sum_{\sigma}\int_{\vec{r},\tau}\bar{\psi}_{\sigma}^{\prime\alpha}(\tau,\vec{r})\left\{ \partial_{\tau}iA^{\alpha}(\tau,\vec{r})\right\} \psi_{\sigma}^{\prime\alpha}(\tau,\vec{r}),
\end{equation}
but it leaves the functional integral measure and the partition function
itself invariant. As a consequence of the latter fact we obtain the
following relation 
\begin{equation}
0=\int\mathcal{D}\mathbf{\Phi}\left(e^{(\bar{\eta},\psi)+(\bar{\psi},\eta)}-e^{-\sum_{\sigma}\int_{\vec{r},\tau}\bar{\psi}_{\sigma}^{\alpha}(\tau,\vec{r})\big\{\partial_{\tau}iA^{\alpha}(\tau,\vec{r})\big\}\psi_{\sigma}^{\alpha}(\tau,\vec{r})+\left(\bar{\eta},e^{iA}\psi\right)+\left(\bar{\psi}e^{-iA},\eta\right)}\right)e^{-S_{R}[\mathbf{\Phi}]+(J,V)}.\label{eq:ExpectationValueWard}
\end{equation}
Considering only an infinitesimal phase transformation this identity
becomes 
\begin{equation}
0=\left\langle \sum_{\alpha=1}^{R}\sum_{\sigma}\int_{\vec{r},\tau}\Big(-\bar{\psi}_{\sigma}^{\alpha}(\tau,\vec{r})\big\{\partial_{\tau}A^{\alpha}(\tau)\big\}\psi_{\sigma}^{\alpha}(\tau,\vec{r})+\bar{\eta}_{\sigma}^{\alpha}(\tau,\vec{r})A^{\alpha}(\tau)\psi_{\sigma}^{\alpha}(\tau,\vec{r})-\bar{\psi}_{\sigma}^{\alpha}(\tau,\vec{r})A^{\alpha}(\tau)\eta_{\sigma}^{\alpha}(\tau,\vec{r})\Big)\right\rangle _{\mathbf{J}}.\label{eq:ExpectationValueWardInfinitesimal}
\end{equation}
The phase field $A^{\alpha}$ may be eliminated entirely by taking
the derivative $\frac{\delta}{\delta A^{\alpha}(\tau)}$. After performing
a Fourier transform we find 
\begin{equation}
0=\left\langle \sum_{\sigma}\int_{\vec{k},\omega}\left(i\nu\bar{\psi}_{\sigma}^{\alpha}(i\omega+i\nu,\vec{k})\psi_{\sigma}^{\alpha}(i\omega,\vec{k})+\bar{\eta}_{\sigma}^{\alpha}(i\omega+i\nu,\vec{k})\psi_{\sigma}^{\alpha}(i\omega,\vec{k})-\bar{\psi}_{\sigma}^{\alpha}(i\omega+i\nu,\vec{k})\eta_{\sigma}^{\alpha}(i\omega,\vec{k})\right)\right\rangle _{\mathbf{J}}.
\end{equation}
Performing the same steps as in the previous section, that is, write
the above equation as a functional differential equation for $\mathcal{Z}_{R}$,
switch to $\mathcal{G}_{R}$ and finally perform a Legendre transform,
we find 
\begin{equation}
\sum_{\sigma}\int_{\vec{k},\omega}\left\{ i\nu\frac{\delta^{2}\mathcal{G}_{R}\left[\mathbf{J}\right]}{\delta\eta_{\sigma}^{\alpha}(i\omega+i\nu,\vec{k})\delta\bar{\eta}_{\sigma}^{\alpha}(i\omega,\vec{k})}+\frac{\delta\Gamma_{R}}{\delta\psi_{\sigma}^{\alpha}(i\omega+i\nu,\vec{k})}\psi_{\sigma}^{\alpha}(i\omega,\vec{k})+\bar{\psi}_{\sigma}^{\alpha}(i\omega+i\nu,\vec{k})\frac{\delta\Gamma_{R}}{\delta\bar{\psi}_{\sigma}^{\alpha}(i\omega,\vec{k})}\right\} =0.\label{eq:MasterWard}
\end{equation}
Once again, the second functional derivative of $\mathcal{G}_{R}$
should be replaced by second functional derivatives of $\Gamma_{R}$.
In analogy to the Schwinger-Dyson equations found above one may obtain
the symmetry relations between different vertex functions by applying
derivatives with respect to the fields.

\noindent Here, we want to obtain a relation between the fermionic
self-energy and the Fermi-Bose three-vertex. To this end, we divide
Eq.~(\ref{eq:MasterWard}) by $i\nu$, sum over the replica index
and apply the derivative $\delta^{2}/\delta\bar{\psi}_{\sigma_{1}}^{\alpha}(i\omega+i\nu,\vec{k})\delta\psi_{\sigma_{2}}^{\alpha}(i\omega,\vec{k})$.
In the limit $\nu\rightarrow0$, we find 
\begin{align}
\frac{\delta^{2}}{\delta\bar{\psi}_{\sigma_{1}}^{\alpha}(i\omega+i\nu,\vec{k})\delta\psi_{\sigma_{2}}^{\alpha}(i\omega,\vec{k})} & \sum_{\beta=1}^{R}\sum_{\sigma}\int_{\vec{k}',\omega'}\frac{\delta^{2}\mathcal{G}_{R}\left[\mathbf{J}\right]}{\delta\eta_{\sigma}^{\beta}(i\omega',\vec{k}')\delta\bar{\eta}_{\sigma}^{\beta}(i\omega',\vec{k}')}\nonumber \\
= & \underset{\nu\rightarrow0}{\lim}\,\frac{1}{i\nu}\left[\frac{\delta^{2}\Gamma_{R}}{\delta\bar{\psi}_{\sigma_{1}}^{\alpha}(i\omega+i\nu,\vec{k})\delta\psi_{\sigma_{2}}^{\alpha}(i\omega+i\nu,\vec{k})}-\frac{\delta^{2}\Gamma_{R}}{\delta\bar{\psi}_{\sigma_{1}}^{\alpha}(i\omega,\vec{k})\delta\psi_{\sigma_{2}}^{\alpha}(i\omega,\vec{k})}\right]+\cdots.\label{eq:DerivativeMasterWard}
\end{align}
The remaining terms, indicated as dots ``$\cdots$'', vanish after
the sources have been set to zero. Next, we need to invoke the Fourier
transformed bosonic Schwinger-Dyson equation (\ref{eq:VertexSchwingerDysonC})
at vanishing boson momentum $\vec{q}=0$ and insert it into the left
hand side of Eq.~(\ref{eq:DerivativeMasterWard}) to replace the
second functional derivative of $\mathcal{G}_{R}$. Finally, we set
the source fields to zero, which yields the Ward-identity presented
in Fig. 3(b) of the main text 
\begin{align}
-\frac{\delta^{3}\Gamma_{R}}{\delta\bar{\psi}_{\sigma_{1}}^{\alpha}(i\omega,\vec{k})\delta\psi_{\sigma_{2}}^{\alpha}(i\omega,\vec{k})\delta V(0)}\bigg|_{\mathbf{\Phi}=\mathbf{\Phi}_{c}} & =\delta_{\sigma_{1},\sigma_{2}}-\underset{\nu\rightarrow0}{\lim}\frac{1}{i\nu}\left[\Sigma_{\sigma_{1}\sigma_{2}}^{\alpha\alpha}(i\omega+i\nu,\vec{k})-\Sigma_{\sigma_{1}\sigma_{2}}^{\alpha\alpha}(i\omega,\vec{k})\right]\nonumber \\
 & =\delta_{\sigma_{1},\sigma_{2}}-\partial_{i\omega}\Sigma_{\sigma_{1}\sigma_{2}}^{\alpha\alpha}(i\omega,\vec{k}).\label{eq:Ward-Id}
\end{align}

\subsection{Replica limit}

\label{sec:ReplicaLimit} To make use of the Ward identity (\ref{eq:Ward-Id})
within the Schwinger-Dyson equation~(\ref{eq:SD_Sigma}), we have
to make the crucial approximation
\begin{equation}
\frac{\delta^{3}\Gamma_{R}}{\delta\bar{\psi}_{\sigma_{1}}^{\alpha}(i\omega,\vec{k})\delta\psi_{\sigma_{2}}^{\alpha}(i\omega,\vec{k}+\vec{q})\delta V(-\vec{q})}\bigg|_{\mathbf{\Phi}=\mathbf{\Phi}_{c}}\quad\longrightarrow\quad\frac{\delta^{3}\Gamma_{R}}{\delta\bar{\psi}_{\sigma_{1}}^{\alpha}(i\omega,\vec{k})\delta\psi_{\sigma_{2}}^{\alpha}(i\omega,\vec{k})\delta V(0)}\bigg|_{\mathbf{\Phi}=\mathbf{\Phi}_{c}},\label{eq:approx}
\end{equation}
whose range of validity has been discussed in the main text. Within
this approximation the self-energy~(\ref{eq:SD_Sigma}) becomes 
\begin{align}
\Sigma_{\sigma_{1}\sigma_{2}}^{\alpha\alpha}\left(i\omega,\vec{k}\right)=\delta_{\sigma_{1},\sigma_{2}}\mathcal{K}(0)\sum_{\beta=1}^{R}\sum_{\sigma}\int_{\vec{k}',\omega'}G_{\sigma\sigma}^{\beta\beta}(i\omega',\vec{k}')+\sum_{\sigma}\int_{\vec{q}}F\left(\vec{q}\right)\left[\delta_{\sigma_{1},\sigma}-\partial_{i\omega}\Sigma_{\sigma_{1}\sigma}^{\alpha\alpha}(i\omega,\vec{k})\right]G_{\sigma\sigma_{2}}^{\alpha\alpha}(i\omega,\vec{k}+\vec{q}).\label{eq:SDW_Sigma}
\end{align}
The physical self-energy is given by the replica limit $\Sigma=\underset{R\rightarrow0}{\mathrm{lim}}\frac{1}{R}\sum_{\alpha=1}^{R}\Sigma^{\alpha\alpha}$.
In this limit the Hartree term vanishes, since it comes with an excess
factor of $R$. (Note that the Hartree self-energy for a single replica
$\alpha$ already contains a summation over a replica index $\beta$,
and thus is proportional to $R$.) Likewise, the bosonic self-energy,
which involves internal fermion loops as well, vanishes, such that
the full bosonic propagator $F(\vec{q})$ is replaced by the bare
propagator $\mathcal{K}(\vec{q})$. Thus, we arrive at Eq. (3)
of the main paper, 
\begin{equation}
\Sigma_{\sigma_{1}\sigma_{2}}\left(i\omega,\vec{k}\right)=\sum_{\sigma}\int_{\vec{q}}\,\mathcal{K}\left(\vec{q}\right)\left[\delta_{\sigma_{1},\sigma}-\partial_{i\omega}\Sigma_{\sigma_{1}\sigma}(i\omega,\vec{k})\right]G_{\sigma\sigma_{2}}(i\omega,\vec{k}+\vec{q}).
\end{equation}

\bibliographystyle{apsrev4-1}
\bibliography{/home/bjoern/Physics/PhD_FU/library}

\end{document}